\documentstyle{aipproc2}

\input epsf

\begin{document}
\title{Inflation and the cosmic microwave background}

\author{Andrew R Liddle}
\address{Astrophysics Group, The Blackett Laboratory, Imperial College,\\
	Prince Consort Road, London SW7 2BZ, Great Britain}

%\lefthead{LEFT head}
%\rig436thead{RIGHT head}

\maketitle

\begin{abstract} 
Various issues concerning the impact of inflationary models on parameter
estimation from the cosmic microwave background are reviewed, with
particular focus on the range of possible outcomes of inflationary
models and on the amount which might be learnt about inflation from the
microwave background.  
\end{abstract}

\section{Introduction}

Inflation \cite{genrefs,LL} maintains its position as the favourite model
for the origin of  structure in the Universe. The reasons are two-fold.
Firstly, the gaussian adiabatic and nearly scale-invariant density
perturbations that the usual models produce currently offer the best
framework in which to interpret observational data on structures in the
Universe. And secondly, it is a extremely simple paradigm within which
to make theoretical predictions; for example the {\sc cmbfast} program
allows the prediction of microwave background anisotropies to better
than one percent accuracy, and as yet calculations in the rival, much
more theoretically challenging, topological defects theories \cite{VS}
lag some way behind \cite{PSTetc}.

However, the inflationary paradigm is quite a broad one, and there
exists a wide range of different implementations of the inflationary
idea. In this article I aim to give a flavour of the complexity of
models which we might one day find ourselves forced to deal with, if we
are to understand structure formation. I will also stress the importance
of these notions for issues such as cosmological parameter estimation
from the cosmic microwave background \cite{parest}, which as we will see
cannot be divorced from the issue of the initial conditions for
structure formation.

In particular, I seek to provide at least partial answers and discussion
centred around the following key questions:
\begin{itemize}
\item What do the simplest models of inflation predict?
\item Is it possible to test the idea of inflation?
\item If the simplest models are right, what do we learn about the 
inflationary mechanism?
\item If the simplest models are not right, what happens then?
\end{itemize}

\section{Cosmological Inflation}

Inflation is defined as a period of accelerated expansion during the
very early stages of the Universe's evolution. However to fix the
physical significance of inflation in your mind, it is better to rewrite
that somewhat, as follows
\begin{equation}
\ddot{a} > 0 \Longleftrightarrow  \frac{{\rm d}}{{\rm d}t} \left( aH
    \right) > 0 \Longleftrightarrow \mbox{The \underline{comoving} 
    Hubble length $H^{-1}/a$ is decreasing.}
\end{equation}
Comoving units are the right ones to use, since to a first approximation
everything, including linear density perturbations, gets dragged along
with the expansion and we are interested in how things evolve relative to that. 
The Hubble length is the key characteristic scale in
the Universe, setting the length scales over which causal processes can
act. What this tells us therefore is that inflation is precisely the
condition that we see a smaller and smaller region of the comoving
Universe as inflation proceeds. In that sense, inflation is rather akin
to `zooming in' on a small region of the initial Universe.

\begin{figure}
\centering
\leavevmode\epsfysize=7cm \epsfbox{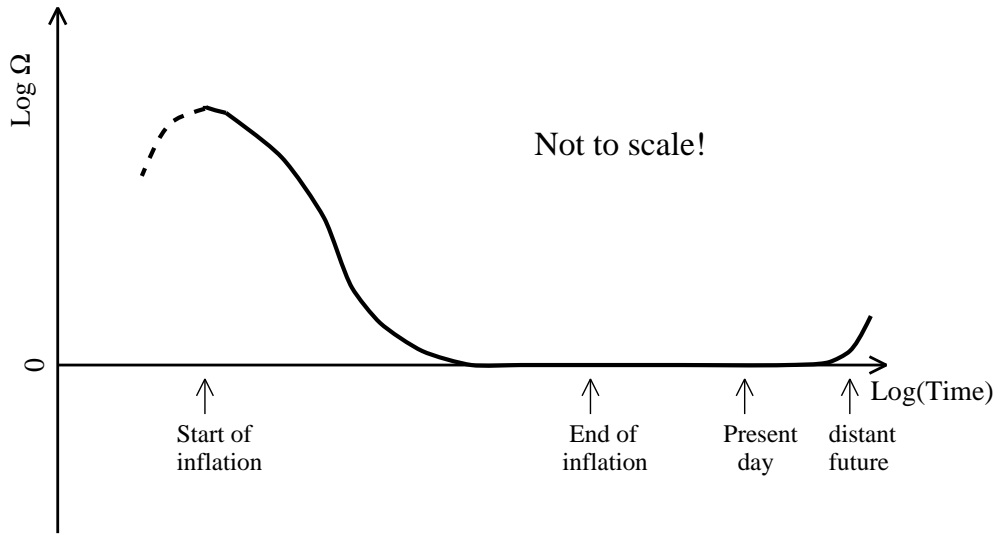}\\ 
\caption[flatness]{\label{f:flatness}A schematic illustration of the
inflationary solution to the flatness problem. Inflation drives the
Universe so close to flatness that only in the extremely distant future is
it possible for the Universe to deviate significantly. [In this figure
$\Lambda$ is assumed zero; otherwise the argument still applies with
$\Omega$ replaced by $\Omega+ \Omega_\Lambda$.]}
\end{figure}

It is well known that inflation solves the classic cosmological
problems. For example, the flatness problem is solved because the
density parameters of matter, $\Omega$, and a possible cosmological
constant, $\Omega_\Lambda$, obey the Friedmann equation
which can be written as
\begin{equation}
\left| \Omega + \Omega_\Lambda - 1 \right| = \frac{|k|}{a^2 H^2} \,.
\end{equation}
By definition, during inflation the right-hand side tends to zero and so
we rapidly approach
\begin{equation}
\Omega + \Omega_\Lambda = 1 \,,
\end{equation}
which is the condition for a spatially-flat Universe. Although after
inflation is over this tells us that the Universe must necessarily be
evolving away from flatness, provided sufficient inflation took place we
would be forced so close to flatness that all the subsequent evolution,
from the end of inflation to the present, would be insufficient to to
move us significantly away from flatness --- see Figure~\ref{f:flatness}.

\begin{figure}[t!]
\centering
\leavevmode\epsfysize=16cm 
\epsfbox{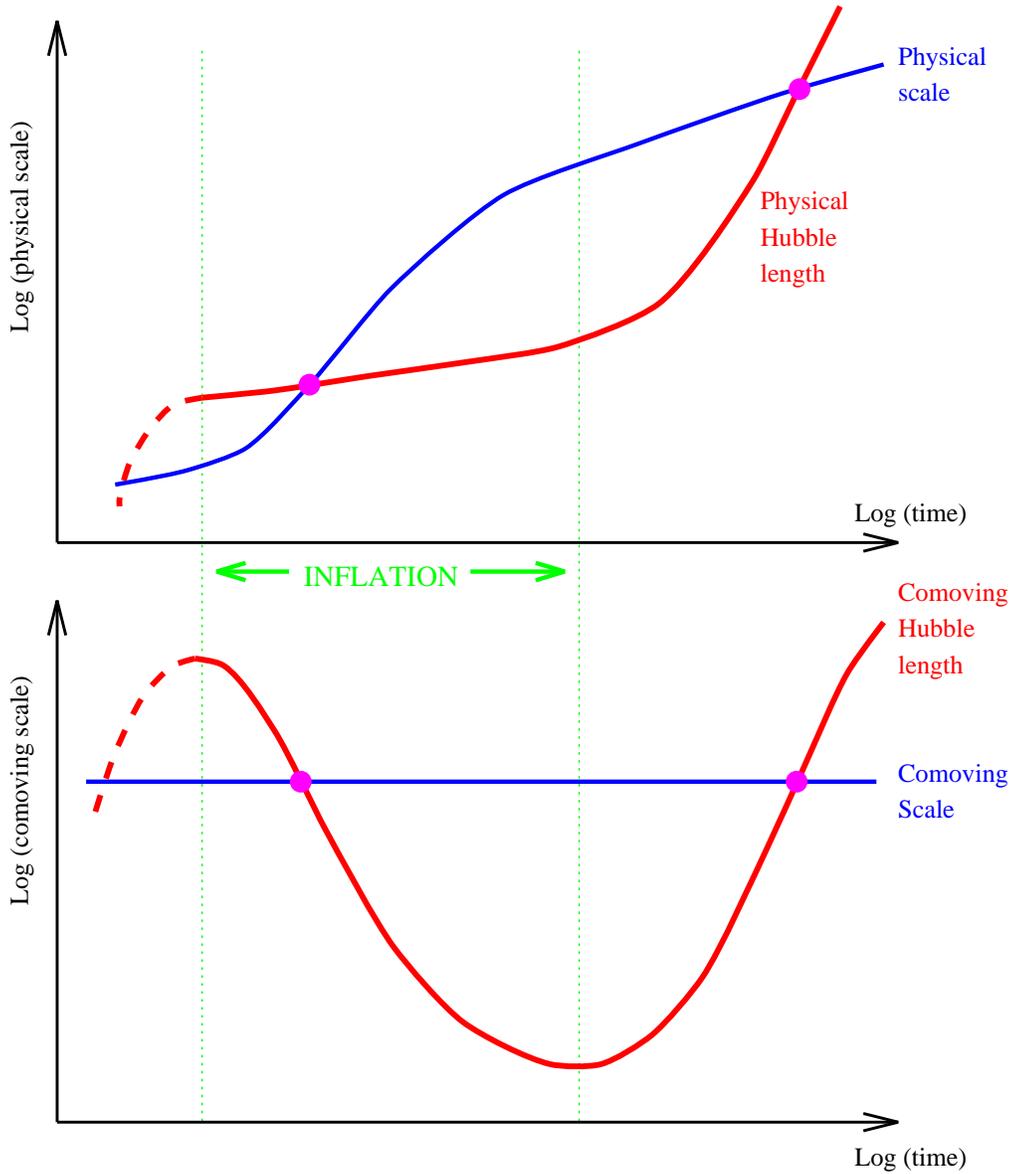}\\ 
\caption[scales]{\label{f:scales} Two equivalent views of how
scales evolve during inflation, as compared to the Hubble length which
sets the scale of causality. The comoving picture is more
transparent, when one recalls that inflation is defined as an epoch
where the comoving Hubble length is decreasing. Scales of interest to us
today begin much smaller than the Hubble length, where quantum
fluctuations are acquired, and by the end of inflation have become much
larger than the Hubble length.}
\end{figure}

While it is an attractive feature that inflation can solve the flatness
and horizon problems, these are `post-dictions' and offer little
prospect of a further test of the paradigm, far less a way to
distinguish between different inflationary models. Consequently,
interest has rightly been refocussed on the fact that the condition for
inflation is precisely that which permits large-scale perturbations to
be generated causally, as it means that scales are `expanded' more
rapidly than the Hubble scale. This is shown in Figure~\ref{f:scales}.
Further, inflation offers a definite mechanism --- that the
perturbations originate as quantum fluctuations --- which in a given
model can be readily calculated \cite{LL}.

\section{Inflationary models {\rm I}}

In the simplest scenarios, which have been widely explored (see
Ref.\cite{LLKCBA} and references therein), inflation is driven by the
potential energy of a single scalar field $\phi$. This potential energy
$V(\phi)$, possibly accompanied by a specific mechanism to end
inflation, is the model input, and physical observables will be
determined by the value of $V$ and its derivatives at the time when
those scales cross outside the Hubble radius during inflation. In the
simplest models this is the only epoch which matters; before then the
scale is well within the Hubble radius and the effect of expansion is
negligible, while afterwards the scale is too large to be influenced by
causal processes. However, we will see later that this is true only of
this simplest class of models.

As far as perturbations are concerned, the simplest models have the
following characteristics.
\begin{enumerate}
\item As there is only one type of matter, isocurvature perturbations
cannot be supported and scalar 
perturbations must be adiabatic. I'll call their spectrum 
$\delta_{{\rm H}}(k)$ (see Ref.~\cite{LL} for a formal definition),
where $k$ is the comoving wavenumber.
\item Gravitational waves are always present at  some level, and I'll
write their spectrum as $A_{{\rm G}}(k)$. In many models, especially
those of the currently-popular hybrid type (see Ref.~\cite{LR} for a
thorough review of inflationary model building) they in fact turn out to
be negligible. 
\item The two spectra can normally be approximated as power-laws
\begin{equation}
\delta_{{\rm H}}^2(k) \propto k^{n-1} \quad ; \quad A_{{\rm G}}^2(k)
	\propto k^{n_{{\rm G}}}
\end{equation}
where $n$ and $n_{{\rm G}}$ are the spectral indices and are readily 
calculated in a given inflation model. However, the power-law expansions are 
only valid if the inflationary potential is flat enough (see later), and if
the available observations are of sufficiently poor quality. High-quality
observations make higher accuracy demands on the description of the
initial power spectrum.
\item The relative importance of density perturbations and gravitational
waves $r$ (measured in terms of their contribution to large-angle microwave 
background anisotropies) is also readily predicted, and equals approximately
$-2\pi n_{{\rm G}}$. In general $n$ and $r$ are independent.
\end{enumerate}

\section{Parameters and their determination}

From reading some parts of the literature, one might form the impression
that the cosmological parameters, such as the Hubble parameter and the
density of the Universe, are directly inscribed on the last-scattering
surface and hence available to be measured in a model-independent way.
Nothing could be further from the truth; the cosmological parameters
govern the {\em dynamics} of perturbations, and a single time-slice in
isolation can say nothing whatsoever about their values. To obtain them,
one needs to understand the perturbations at some other epoch, and since
this is not accessible to direct observation this initial form must be
fixed by assumption. Since there is no unique model supplying these
initial conditions, they must be parametrized in some way and those
parameters added into the melting pot. The various parameters can be
divided into the cosmological parameters (whose role is to describe the
present state of the Universe) and the inflationary parameters,
describing the initial conditions. A typical set might be

\begin{center}
{\large Cosmological parameters}
\begin{quote}
\begin{description}
\item[$h$:~~~] The Hubble parameter.
\item[$\Omega_0$:~~] The density of matter in the Universe.
\item[$\Lambda$:~~~] A cosmological constant, if present.
\item[$\Omega_{{\rm B}}$:~~] The amount of baryonic matter.
\item[$\Omega_{{\rm HDM}}$:] The amount of hot dark matter, if present.\\
\hspace*{0.1cm}(It is assumed that there is always some cold dark matter.)
\item[$\tau$:~~~] The optical depth to the last scattering surface
caused by reionization.
\end{description}
\end{quote}

{\large Inflationary parameters}
\begin{quote}
\begin{description}
\item[$\delta_{{\rm H}}$:~] The overall amplitude of density perturbations.
\item[$n$:~~] The spectral index of the density perturbations.
\item[$r$:~~] The effect of gravitational waves on COBE.
\end{description}
\end{quote}
\end{center}

Amongst those cosmological parameters, some might be thrown out by
assumption; for example we may restrict ourselves to a flat Universe
without a hot dark matter component. Typically authors have considered
fairly wide cosmological parameter sets. However on the inflationary
side most efforts have been very minimal, often with just $\delta_{{\rm
H}}$ and $n$ considered. While it might be highly desirable to imagine
that the initial conditions can be parametrized by so few parameters,
and while indeed that may well in the end turn out to be true, it needs
to be acknowledged that less kind circumstances might also prevail, as
described in the next section.

Having decided your parameters, you can do two things. The first, which
has received quite a bit of attention, is to make estimates of how
accurately one can expect to measure said parameters, under the
assumption that they provide a sufficient set and for some chosen
experimental configuration \cite{parest}. An example will be given next
section. The second thing is to actually go ahead and try and estimate
the parameters from some data. This has begun to be popular
\cite{paramest}, though so far the only attempt to cover enough
parameters that one can argue they are sufficient is a recent attempt by
Tegmark \cite{hercules} to constrain a nine-parameter space based on the
simplest inflation models. Current data is not really up to the task of
seriously limiting such a wide parameter volume, but hopefully that will
change in the medium future.

\section{Inflationary models {\rm II}}

The simplest inflationary models are the ones most commonly described.
But there are several other possibilities already in the literature,
which would impact on the simple picture outlined so far.

\begin{itemize}
\item {\it There may be only one field, but with a potential 
energy $V(\phi)$ which is less flat than we might like.}\\
Then the power-law approximation might break down. The simplest
manifestation of this is for the spectral index $n$ to become 
scale-dependent, but perhaps amenable to a perturbative approach as
outlined later in this section. 
More drastic is the case of `designer' power spectra,
where the potential has sharp features leading to similarly sharp
features in the spectra. This would need to be dealt with on a
model-by-model basis.
\item {\it There may be more than one field.}\\
Then the adiabatic perturbations might be accompanied by isocurvature 
perturbations. Superhorizon adiabatic perturbations are then no longer 
constant \cite{modes}. We cannot necessarily ignore details of physics
while modes are  superhorizon, e.g.~the (p)reheating epoch which brings
inflation to an end.
\item {\it Quantum tunnelling effects might be important.}\\
Then the Universe might be open instead of flat \cite{openinf}.
\end{itemize}
I will briefly look at the possible impact of the first two of these.

\subsection{Scale-dependent spectral index}

The better the quality of observations available, the more likely it is
that the power-law approximation is inadequate. It is in fact the
beginning of a Taylor expansion of the power spectrum of the form \cite{cgl}
\begin{equation}
\ln \delta_{{\rm H}}^2(k) = \ln \delta_{{\rm H}}(k_0) + (n-1) \ln
    \frac{k}{k_0} + \frac{1}{2} \left. \frac{dn}{d \ln k} \right|_{k_0} \,
    \ln^2 \frac{k}{k_0} + \cdots \,,
\end{equation}
where only the first two terms are kept. The expansion scale $k_0$ is
arbitrary, but a specific choice is in fact favoured in that it can be
chosen so that the estimated errors on $n$ and on $dn/d\ln k$ are
uncorrelated. 

A perturbative approach is therefore suggested where we add further
terms to this series (i.e.~adding the coefficients to the parameter
estimation process) until a satisfactory fit is achieved. In a given
inflation model these coefficients are as easily estimated as $n$
itself, and indeed some inflation models, especially the so-called
`running mass' models \cite{runmass}, can indeed give an effect large
enough to be detectable \cite{cgl}.

When more parameters are added, in principle the error on {\em all}
parameters is increased. We computed the effect in Ref.~\cite{cgl},
assuming a version of the cold dark matter model and observations by a
version of the Planck satellite including polarized detectors, and 
the table shows how the uncertainties change as extra derivatives of $n$
are added. In fact the picture is quite encouraging. The uncertainties
on the cosmological parameters hardly increase at all, and even
the uncertainty on the spectral index itself suffers little as extra
derivatives are incorporated. It is quite possible therefore that in
these models one might get extra information on the inflationary model,
through determination of the higher derivatives, while suffering little
cost elsewhere.

\begin{table*}
\caption[caption]{Estimated uncertainty in parameter estimation from 
measurements by a version of the Planck satellite with polarized detectors. Full 
details in Ref.~\cite{cgl}.}
\begin{tabular}{|l|llll|} 
\hline
&&&&\\
Parameter &\hspace{0.5cm}& \multicolumn{3}{c|}{Planck with polarization} \\
\hline 
&&&&\\
$\delta \Omega_{{\rm b}} h^2 /\Omega_{{\rm b}} h^2$ & &
 $  0.007$ & $  0.008$ & $0.009$ \\
$\delta h/h$ & & $0.01$ & $0.01$ & $ 0.01$ \\
$\delta \Omega_{{\rm \Lambda}} h^2 /h^2$ & & $0.04$ & $0.04$ & $ 0.05$ \\
$\delta \tau$ & & $ 0.002$ & $ 0.002$ & $ 0.002$\\ 
&&&&\\
$\delta n$ & & $ 0.004$ & $0.004$ & $0.006$ \\
$\delta r$ & & $ 0.04$ & $0.04$ & $0.04$ \\
$dn/d\ln k$ & & $-$ & $  0.009$ & $0.01$ \\
$d^{2}n/d(\ln k)^2$ & & $-$ & $-$ & $0.02$ \\
&&&&\\
\hline
\end{tabular}
\end{table*}

\subsection{Isocurvature models}

If there are multiple scalar fields, then we may find isocurvature 
perturbations as well as adiabatic ones. This represents a significant
complication, because adiabatic and isocurvature modes can source each
other even on superhorizon scales if there are interactions between the
field or fluid components. Predicting the perturbations in such models
presents no problem of principle, but extra information might need to be
provided (details of the reheating period ending inflation, for
example), whereas in the single field model causality provides a direct
bridge from the inflationary era until the scales re-enter the horizon. In 
general
very complicated post-inflation processing between isocurvature and
adiabatic modes might occur \cite{Hu}.

One type of model aims to create only isocurvature perturbations
\cite{iso}. The basic idea is that during inflation, the field which
later becomes the CDM already exists and experiences perturbations via
the usual quantum fluctuation mechanism. When it later becomes
dynamically significant, we  have an isocurvature perturbation between
CDM and the rest. 

Alternatively, there may be a mixture of the two. Low-level isocurvature
contamination can impact on parameter estimation, since it will
contribute either as a detectable component requiring extra parameters
for its description, or as an extra undetectable noise contribution.

Inflationary models of these types have to be compared to observational 
data on a model-by-model basis. This potentially leads to a wide range
of predictions. There is however one saving grace, which is that at
least all these models share one characteristic which seems unavoidable,
being the existence of a peak structure in the microwave anisotropy
spectrum \cite{huwhite}. This appears to be a generic prediction, as it
follows from a simple argument.
Because perturbations are established well outside the horizon, they are 
observed entirely in the growing mode. This necessarily leads to a phase
coherence and ultimately to an acoustic peak structure.

The conclusion therefore is that if the observed spectrum fails to
feature a series of peaks when measured at high resolution, then
inflation cannot be the sole source of perturbations originating
structure. Note that phrase carefully however to avoid confusion. If a
peak structure is not observed, that doesn't completely rule out
inflation, because it may partly contribute to the anisotropies, an
example being the mixed inflation plus defects scenario where the peak
structure vanishes if the defects are sufficiently dominant \cite{chm}. Further, 
if
the peak structure {\em is} observed, that does not `prove' inflation.
It would support inflation as it would have passed a further
observational hurdle, but other mechanisms may be able to produce
similar structures.
 
\section{Conclusions}

The main point I have aimed to stress in this article is that the
inflationary paradigm is a broad one, realizable in a number of ways.
The inflationary input is crucial to current views on parameter
estimation, not just of the inflationary parameters but the cosmological
ons too. If the simplest inflationary models prove correct, all will be
well but we will learn only a limited amount about the inflationary
mechanism. More desirable would be something a little more complicated,
allowing us to extract further information about inflation at only a very
modest cost to the cosmological parameters. But we had better hope that
the model is not so complicated that it requires so many additional
parameters as to have a serious effect on our ability to fit the
observational data, or even worse a model which defies parametrization
at all. Fortunately, all the current indications are favourable.

\section*{Acknowledgments}

I thank Ed Copeland and Ian Grivell for collaboration on some of the work 
described in this article.

\end{document}